\documentclass[aps,prl,superscriptaddress,amsmath,amssymb,twocolumn]{revtex4-1}
\usepackage{graphicx,float} 
\usepackage{braket}
\usepackage[utf8]{inputenc}
\usepackage{dcolumn}
\usepackage{xfrac}
\usepackage{hyperref}
\usepackage{bm}
\usepackage{mhchem}

\newcommand{\di}{i} % default math "i"

\begin{document}

\title{Angular Momentum of Fission Fragments from Microscopic Theory}

\author{Petar Marevi\'c}
\affiliation{Nuclear and Chemical Sciences Division, Lawrence Livermore National Laboratory, Livermore, CA
94551, USA}
\author{Nicolas Schunck}
\affiliation{Nuclear and Chemical Sciences Division, Lawrence Livermore National Laboratory, Livermore, CA
94551, USA}
\author{J\o rgen Randrup}
\affiliation{Nuclear Science Division, Lawrence Berkeley National 
Laboratory, Berkeley, CA 94720, USA}
\author{Ramona Vogt}
\affiliation{Nuclear and Chemical Sciences Division, Lawrence Livermore National Laboratory, Livermore, CA
94551, USA}
\affiliation{Physics and Astronomy Department, University of California, Davis, CA 95616, USA}

\begin{abstract}  
During nuclear fission, a heavy nucleus splits into two rotating fragments.
The associated angular momentum is large, yet the mechanism
of its generation and its dependence on the mass of fragments
remain poorly understood.
In this Letter, 
we provide the first microscopic calculations of 
angular momentum distributions in fission fragments 
for a wide range of fragment masses. For the benchmark 
case of $^{239}$Pu($n_{\text{th}}$,f), we find that the angular momentum of the 
fragments is largely determined by the nuclear shell structure
and deformation, and that the heavy fragments therefore typically carry 
less angular momentum than their light partners. We use the fission model
$\tt{FREYA}$ to simulate the emission of 
neutrons and photons from the fragments. The dependence
of the angular momenta on fragment mass after the emission of neutrons
and statistical photons is linear for the heavy fragments and either constant
or weakly linear for the light fragments, consistent 
with the universal sawtooth pattern
suggested by recent experimental data.
Finally, we observe that using microscopic
angular momentum distributions modifies the number of emitted photons significantly.
\end{abstract}
 
\date{\today}

\maketitle

%\thispagestyle{fancy}
%\fancyhead[R]{LLNL-JRNL-818187}

% Introduction and the importance of angular momentum
{\it Introduction. ---} Nuclear fission plays an essential
role in fundamental and applied science, but even eighty
years after its discovery \cite{hahn1939,meitner1939}
the microscopic foundation of the phenomenon is yet to be fully
understood \cite{krappe2012,bender2020}.
In a simple picture, fission can be viewed as the shape
evolution of a charged quantum liquid drop that gradually deforms
while surmounting multiple potential barriers and finally assumes
an extremely deformed shape at scission \cite{bohr1939}.
% 'Fissile' has a very precise meaning
Fissionable nuclei predominantly split into two primary fragments that 
move apart due to the mutual Coulomb repulsion. 
Simultaneously, primary fragments relax toward their equilibrium shapes 
and then deexcite by sequential emission of neutrons and photons.
The angular momentum (AM) of fission fragments (FFs) influences
this deexcitation process, causing the neutron emission to be anisotropic and
affecting the number of emitted photons
\cite{bowman1962,gavron1976,pringle1975,vogt2017,vogt2021}.
However, the mechanism of the AM
generation and its dependence on the mass of fragments
are still poorly understood.
Very recently, a sawtooth-like mass dependence
of the AM suggested in earlier experiments \cite{naik2005} was measured across
three different fission reactions \cite{wilson2021}.
The observed lack of correlation between the 
AM of the fragment partners was interpreted
as a proof of its post-scission origin \cite{wilson2021}, 
although it was immediately pointed out that
this observation could be 
explained by a pre-scission mechanism as well \cite{randrup2021}. 
A full understanding of the AM of FFs will eventually
need to be rooted in a microscopic framework where the fission phenomenon
emerges from internucleon forces
and quantum many-body physics.
However, such a framework
is markedly missing, and the most widely used 
fission models still rely on 
AM distributions that are
obtained from either statistical or semi-classical methods
\cite{vogt2021,verbeke2018,randrup2014,talou2011,becker2013,stetcu2016}. 

% About DFT 
Currently, nuclear density functional theory (DFT) \cite{schunck2019} is
the only fully quantum-mechanical approach capable of 
describing many facets of fission
\cite{schunck2016,bender2020}. In particular, DFT models 
were successfully employed in studies of spontaneous 
fission half-lives \cite{baran2015}, FF mass and charge 
distributions
\cite{goutte2004,regnier2016a,tao2017,regnier2019,verriere2019,
zhao2019,zhao2019a}, and energy sharing among the FFs
\cite{younes2011,simenel2014,bulgac2016,bulgac2019}. 
There were a few attempts at describing the AM of FFs within the DFT framework, 
but they either relied on strong and restrictive approximations \cite{bonneau2007,bertsch2019} 
or were limited to a single pair of fragments only \cite{bulgac2020}.
Consequently, a quantitatively robust and predictive 
framework for computing AM distributions over the wide range of FF masses 
that is typically observed in experiments \cite{wilson2021,schillebeeckx1992,nishio1995} 
is still missing.

% Summary of what the letter brings
In this Letter, we report the first microscopic
calculations of AM distributions in FFs for a wide range of fragment masses. 
Starting 
from a large set of extremely deformed
scission configurations,
projection techniques are employed to extract the AM distributions 
of $24$ fragment pairs in $^{239}$Pu($n_{\text{th}}$,f). We find
that the AM of FFs is largely determined by the underlying shell
structure and deformation of fragments at scission. As a result, 
the heavy FFs typically carry less angular
momentum than their light partners. 
By adapting the fission simulation model $\mathtt{FREYA}$ 
to use the microscopic AM distributions, we are able to 
simulate the emission of neutrons
and photons from primary FFs. We find that, after the 
emission of neutrons and statistical photons, the dependence
of the AM on the mass of FFs is linear for the heavy fragments 
and either constant
or weakly linear for the light fragments. 
This result is consistent with a universal sawtooth
pattern proposed in recent experiments.
Finally, we assess the impact of microscopic
AM distributions on total neutron 
and photon multiplicities and find that
the latter are significantly
modified.

% Theoretical framework
{\it Method. ---} A set of scission
configurations is first determined by solving the Hartree-Fock-Bogoliubov (HFB)
equations with the $\mathtt{HFBTHO}$ package \cite{perez2017}, 
using the SkM* parameterization of the Skyrme energy 
functional \cite{bartel1982} and a mixed volume-surface 
contact pairing force \cite{dobaczewski2002}. Constraints are imposed on the values of the axially-symmetric 
quadrupole $(q_{20})$ and octupole $(q_{30})$ moments \cite{dobaczewski2004}, 
corresponding to the elongation and the reflection asymmetry 
of the nuclear shape, respectively. In addition,  we constrain the expectation value of the neck operator 
$(q_N)$, which estimates the number of nucleons in a 
thin neck connecting the two fragments \cite{younes2009}. 
This approach enables us to explore a three-dimensional 
$\bm{q} \equiv (q_{20}, q_{30}, q_N)$ hypersurface in collective space
and to generate a large set of scission 
configurations \cite{schunck2014}. A total of $1545$ 
configurations $\ket{\Phi(\bm{q})}$ in $^{240}$Pu are considered, with neck 
values $1 \leq q_N \leq 3$ and a wide range of 
quadrupole and octupole deformations;
see the Supplemental Material \cite{supplemental_material} 
for more details on technical aspects of the HFB calculation
and on the properties of scission configurations.

The scission configurations have axially-symmetric
densities that are dumbbell-shaped and can readily be
divided into heavy ($z<z_N$) and light ($z>z_N$) FFs,
where $z_N$ locates the minimum of the density profile.
By adapting standard symmetry restoration techniques
\cite{ring2004,sheikh2019,bally2021} to the case of FFs,
AM distributions of the heavy ($F=H$) and the light fragment ($F=L$) 
for each configuration $\ket{\Phi(\bm{q})}$ can be calculated as
\begin{equation}
|a_{J}^{F}(\bm{q})|^2 = \int_{\beta} \braket{\Phi(\bm{q})|
\hat{R}_y^{F}(\beta)|\Phi(\bm{q})},
\label{eq:distributions}
\end{equation}
where $\int_{\beta} \equiv (J+{1\over2}) \int_0^{\pi} \,d\beta
\sin\beta \, d^{J*}_{00}(\beta)$ denotes integration over the 
orientation angle $\beta$ with Wigner matrix elements
$d^{J}_{00}(\beta) = P_{J}(\cos\beta)$ \cite{varshalovich1988}
($P_J$ is the Legendre polynomial of order $J$) as weights,
and $\hat{R}_y^{F}(\beta) = 
\exp(-\di \beta \hat{J}_y^{F})$ is the rotation operator for 
fragments. The angular momentum operators $\hat{J}_y^{F}$ have support
within the spatial region $\mathcal{S}^{F}$ containing each fragment
\cite{sekizawa2014,sekizawa2017}. They are computed from 
the associated kernels,
%\fancyhead[R]{}
\begin{equation}
J_y^{F}(\bm{r},\sigma) = \Theta^{F*}(z-z_N) J_y(\bm{r},\sigma)
\Theta^{F}(z-z_N),
\label{eq:kernel}
\end{equation}
where  $J_y(\bm{r},\sigma) = L_y(\bm{r}) + 
S_y(\sigma)$ corresponds to the usual angular momentum operator that depends 
on the spatial coordinates $\bm{r} \equiv (r_{\perp}, \phi, z)$ 
and the spin coordinate $\sigma$, $\Theta^{H}(z-z_N) = 1 - 
\mathcal{H}(z-z_N)$, $\Theta^{L}(z-z_N) = \mathcal{H}(z-z_N)$, 
and $\mathcal{H}(z)$ is the Heaviside step function 
\cite{abramowitz1965}. The center of mass of each 
fragment is located at $\bm{r}_{\text{CM}}^{F} = 
(0, 0, z_{\text{CM}}^{F})$. Therefore, we take $\bm{r} 
\rightarrow \bm{r} - \bm{r}_{\text{CM}}^{F}$ in 
Eq.~(\ref{eq:kernel}) to determine the angular momentum with respect 
to the center of mass of each fragment. To ensure a proper 
convergence of integrals in Eq.~(\ref{eq:distributions}) for 
all $J$ values, $N_{\beta}=60$ orientation angles are taken into 
account. Note that the configurations $\ket{\Phi(\bm{q})}$ are expanded
in a basis that is not closed under rotation
\cite{robledo1994}. Therefore, we must employ the recently introduced 
technique of symmetry restoration in incomplete bases 
\cite{marevic2020} to evaluate Eq.~(\ref{eq:distributions}).

The number of nucleons in FFs can be obtained 
as the integral of the total density over the subspace $\mathcal{S}^{F}$ containing 
each fragment. Thus, this calculation represents 
a mapping of a set of collective variables $\bm{q}$ in the 
combined nuclear system ($Z, A$) onto a set of charges and masses
($Z_{F}(\bm{q}), A_{F}(\bm{q})$) in the two fragments. These charges and masses
satisfy $Z_{H}(\bm{q})+Z_{L}(\bm{q})=Z$ and $A_{H}(\bm{q})+A_{L}(\bm{q})=A$
but are generally 
not integers. In principle, one should combine the 
outlined method with the particle number projection (PNP) in FFs
\cite{scamps2017,verriere2019,bulgac2019a}, a formidable task 
that is yet to be undertaken. To extract the desired quantities 
for integer numbers of nucleons in each fragment, we instead 
perform a Gaussian Process interpolation \cite{rasmussen2006,scikit-learn}
over a subset of 
scission configurations with both the proton and the neutron 
number in the vicinity of the target values.
Therefore, the obtained distributions $|a_J|^2$ are complemented by the 
corresponding confidence intervals $\sigma(|a_J|^2)$. Finally, the present model is limited to states that obey 
the natural
spin-parity rule, $(-1)^J= \pi$. Consequently, we cannot
extract odd $J$
if $q_{30}^F \rightarrow 0$ \cite{supplemental_material}. To account for these rare cases, we 
solve Eq.~(\ref{eq:distributions}) for even $J$, interpolate 
for odd $J$, and normalize the entire distribution. This procedure smoothens the distributions of $q_{30}^F \rightarrow 0$
configurations while having a negligible effect on the average 
angular momentum they carry. More details on the
entire interpolation procedure can be found in the Supplemental Material
\cite{supplemental_material}. 

%%%%%%%%%%%%%%%%%%%%%%%%%%%%%%%%%%%%%%%%%%%%%%%%
% Results
{\it Results. ---} Starting from a set of $1545$ scission 
configurations, we determined the AM distributions $|a_J^F|^2$ of 
$24$ fragment pairs in $^{239}$Pu($n_{\text{th}}$,f) within the 
mass range $126 \leq A_{H} \leq 150~(90 \leq A_L \leq 114)$, covering more than $95\%$ of 
measured mass yields
 \cite{schillebeeckx1992,nishio1995}.
With a few exceptions, the employed 
collective space $\bm{q}$ allows the extraction of only 
one $Z$ value per each $A$ (see Figs.~1a and 1b of Ref.~\cite{supplemental_material}).
A broader charge dispersion could be obtained
by performing the additional PNP
in FFs.
Furthermore, this initial set comprised a wide range of 
quadrupole and octupole deformation parameters of the FFs,
$\beta_2^F(\bm{q}) \leq 0.7$ and $|\beta_3^F (\bm{q})| 
\leq 0.5$, where 
$\beta^F_\lambda = (4\pi)/(3A_FR_F^\lambda)q_{\lambda 0}^F$,
$q_{\lambda 0}^F$ are the multipole moments \cite{scamps2019} of the FFs,
and $R_F=1.2 A_F^{1/3}$ fm. The same Gaussian process 
procedure was used to extract these parameters for the 
integer-valued FFs \cite{supplemental_material}. 
We obtained FFs with a wide range of quadrupole deformations
and confirmed the findings of Refs.~\cite{bulgac2016,scamps2019a} that FFs
are octupole-deformed at scission. 

\begin{figure}[!ht]
\includegraphics[width=0.49\textwidth]{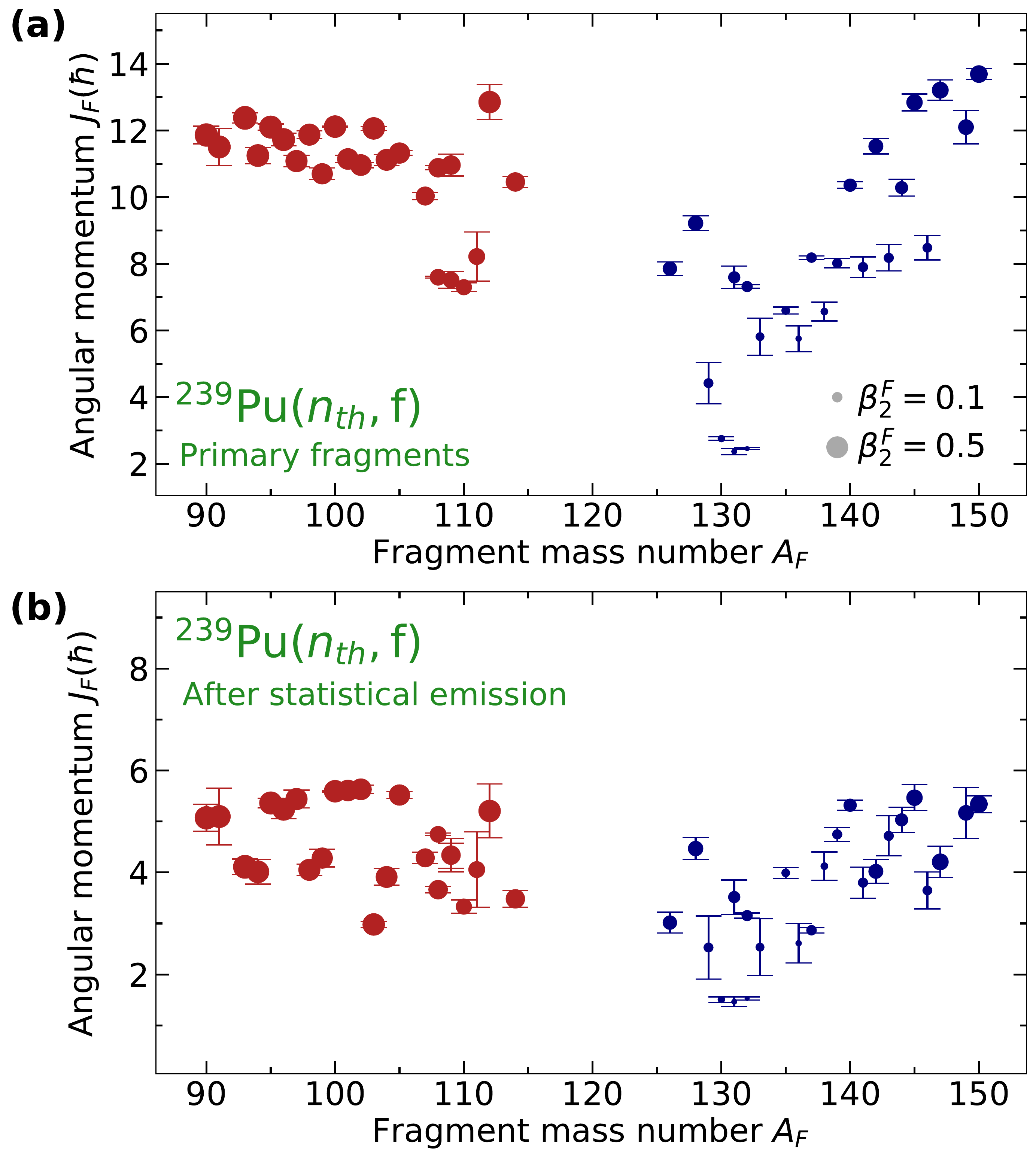}
\caption{Average angular momentum $J_F$ of the heavy (blue) and light (red) fission fragments in
$^{239}$Pu($n_{\text{th}}$,f) as a function
of their mass number $A_F$. (a): Average AM of primary fission fragments
calculated with microscopic theory.
(b): Average AM of the fragments after
the emission of
neutrons and statistical photons was
simulated with $\tt{FREYA}$. In both panels, the area of the circles is
proportional to the quadrupole deformation parameter 
$\beta_2^F$ extracted for the fragments at scission. 
Note that the legend is illustrative and 
does not represent two discrete cases. 
The error bars in both panels represent the $\pm 2\sigma_F^{\text{GP}}$ 
confidence intervals stemming from the Gaussian process interpolation 
for integer numbers of nucleons in FFs, where $\sigma_F^{\text{GP}}  =
\sum_J |\frac{\partial J_F (J_F+1)}{\partial |a_J|^2} \sigma(|a_J|^2)|/(2J_F+1)$.}
\label{fig:average_AM}
\end{figure}

Fig.~\ref{fig:average_AM}a shows the
average angular momentum of the heavy and the light primary FFs 
as a function of their mass number $A_F$. The average values $J_F$ are obtained from
$J_F (J_F+1) = \sum_{J=0}^{30} J(J+1)|a_J^F|^2$. The corresponding error bars
stem from the Gaussian process interpolation
for integer numbers of nucleons in FFs and they
do not represent the total theoretical
uncertainty. In particular, another important source of  uncertainty 
is the choice of the energy density functional; the associated uncertainty 
is likely of the order of $1 \hbar$ \cite{bulgac2020}
and may be mass dependent.
The heavy FFs display a wide range of 
angular momenta and quadrupole deformations that appear to be strongly 
correlated. The smallest values are found in the region of
the doubly magic $\ce{^{132}_{50}}$Sn nucleus 
while rare-earth isotopes are highly deformed and on average carry substantial 
angular momentum. The average values around 
the most likely fragmentation ($A_H$:$A_L \approx 136$:$104$) are in the range $5-9\,\hbar$. 
On the other hand, the light 
FFs are significantly more deformed and therefore on average 
carry more angular momentum, typically $10-13\,\hbar$. 
A similar conclusion, but for the most likely fission
fragmentation only, was very recently obtained within
the framework of time-dependent DFT \cite{bulgac2020}.
These results are at odds with the outputs of the widely-used 
phenomenological models such as $\mathtt{FREYA}$ \cite{randrup2014,verbeke2018,vogt2021}
and $\mathtt{CGMF}$ \cite{talou2011,becker2013,
stetcu2016} where the AM of the FFs is calculated using
generic moments of inertia, $\mathcal{I}_F$,
that omit important structure and deformation effects. 
These generic $\mathcal{I}_F$ increase with mass, 
giving
$\mathcal{I}_H > \mathcal{I}_L$, leading
to a higher AM for the
heavy fragment, $J_H > J_L$.
In a separate publication, two
of us show 
that including 
deformation effects in a phenomenological
model indeed reverses this trend \cite{randrup2021}.
In addition to enabling 
asymmetric fragmentations by stabilizing octupole deformations
\cite{scamps2019} and modifying the structure of fission barriers \cite{brack1972}, 
we demonstrate here that the shell effects \emph{at scission}
also significantly influence the angular momentum of the fission fragments.

\begin{figure*}[!htb]
\includegraphics[width=0.99\textwidth]{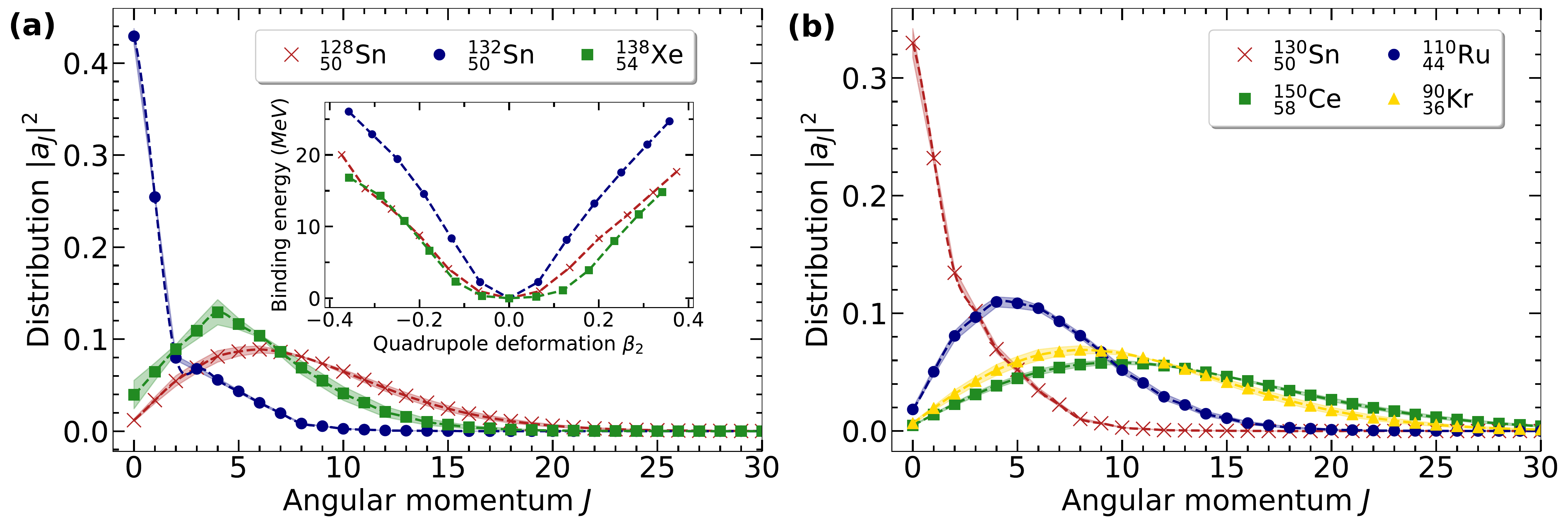}
\caption{Angular momentum distributions of  
primary fission fragments in 
$^{239}$Pu($n_{\text{th}}$,f). (a): The distributions for 
three heavy fragments in the vicinity of the $Z=50$ and $N=82$ magic 
numbers. The inset shows
the binding energy of these nuclei 
(normalized to the corresponding absolute minima)
as a function of the quadrupole deformation parameter $\beta_2$,
calculated with the HFB model.
(b): The distributions for the partner fragments
from two different mass divisions. In both panels, 
symbols mark the calculated values and
the lines are obtained by a simple spline interpolation.
The shaded areas show the $\pm 2\sigma(|a_J|^2)$ confidence intervals 
from the Gaussian process interpolation for integer numbers
of nucleons in FFs.
}
\label{fig:fragmentations}
\end{figure*}

To shed more light on the effect of the underlying shell 
structure on the angular momentum of primary FFs, we show in Fig.~\ref{fig:fragmentations}a
the AM distributions 
of three fragments in the vicinity of the $Z=50$ and $N=82$ magic 
numbers. Four neutrons away from the double shell closure, $\ce{^{128}_{50}}$Sn
has an average
AM of $J_H(\sigma_H^{\text{GP}}) \approx 9.2(0.2)\,\hbar$. 
The situation drastically changes at $N=82$ where $J_H \approx 2.5(0.1)\,\hbar$
for $\ce{^{132}_{50}}$Sn. Adding four more nucleons leads to $J_H \approx 6.6(0.3)\,\hbar$
in $\ce{^{138}_{54}}$Xe.
The most advanced microscopic models so far relied on the assumption that FFs
can be represented by deformed, isolated nuclei with the same number of 
nucleons \cite{bonneau2007,bertsch2019}.
As shown in the inset of Fig.~\ref{fig:fragmentations}a, these three nuclei
are all spherical in their ground states and are thus beyond the scope of models 
that assume the existence of rigidly deformed equilibria \cite{bonneau2007}. 
Considering instead a range of arbitrary deformations around the equilibrium 
configuration \cite{bertsch2019} does not yield unique values of angular momenta
and consequently provides a very limited predictive capability.
In contrast, the present approach considers FFs within the
actual combined nuclear system at scission. Consequently,
deformations of FFs are automatically
determined by the variational principle and are 
generally different from their equilibrium deformations. 
This imparts a predictive power to the model
and can naturally lead to different AM distributions for FFs with similar equilibrium deformations.

Furthermore, we employed the fission model $\mathtt{FREYA}$ 
to simulate the emission of neutrons
and photons from primary FFs and
assess the impact of the microscopic AM distributions on neutron 
and photon multiplicities. $\mathtt{FREYA}$ \cite{randrup2014,verbeke2018,vogt2021} 
is a Monte Carlo model that generates large samples of complete 
fission events, providing the full kinematic information
for the two product nuclei and the emitted neutrons and photons in each event.
Inputs to $\mathtt{FREYA}$ are primarily based on experimental 
data, although a certain degree of modeling is necessary.
%%%
In particular, $\mathtt{FREYA}$ obtains the AM of FFs from a statistical
sampling of the dinuclear wriggling and bending modes, 
employing an adjustable spin temperature $T_S$ \cite{vogt2017,vogt2021}.
For each fragmentation considered, it is possible to
adjust $T_S$ such that the AM distribution generated by $\mathtt{FREYA}$ closely
matches the one extracted  microscopically.
Consequently, we can
determine the neutron and photon observables
resulting from the microscopic AM distributions. For each fragmentation considered, we simulated
two hundred thousand fission events and calculated the average angular momenta
after the emission of neutrons and statistical photons (Fig.~\ref{fig:average_AM}b).
The neutron evaporation reduces the 
average AM only marginally (less than $0.4 \hbar$ in each FF),
while statistical photons typically remove several 
units of $\hbar$. Since different FFs can have markedly
different AM distributions (Fig.~\ref{fig:fragmentations}b), 
the angular momentum is not reduced
by a constant value. For example,
the reduction for the pair of fragments $\ce{^{130}_{50}}$Sn and $\ce{^{110}_{44}}$Ru 
is $\approx 1 \hbar$ and $\approx 4 \hbar$, respectively. On the other hand, 
the AM distributions of $\ce{^{150}_{58}}$Ce and $\ce{^{90}_{36}}$Kr are more 
similar and statistical
emission reduces their large angular momenta
acquired at scission by
$\approx 8 \hbar$ and $\approx 7 \hbar$, respectively.
The emission of statistical
photons ceases when a nucleus reaches the yrast line and deexcitation to the ground
state then proceeds mostly through the emission
of quadrupole photons. 
We compared the average number of
neutrons and photons (multiplicities) emitted from each FF during
the entire deexcitation process 
calculated with (i) default $\tt{FREYA}$ AM distributions and 
(ii) microscopic AM distributions.
The neutron multiplicities depend 
rather weakly on AM distributions, in agreement with the previously 
reported observation that the rotation of FFs influences the 
direction of emitted neutrons but not their number \cite{vogt2021}. 
On the other hand, the photon multiplicities are significantly modified, 
as shown in Table~\ref{tab:multiplicities}. In particular, 
the microscopic calculations yield lower $J_H$ in the vicinity of the double
shell closure, leading to lower multiplicities. The trend reverses at 
larger and smaller $A_H$, where larger 
deformations of FFs
lead to larger multiplicities. 
Extending the present model to other
fragmentations will enable a full-fledged $\mathtt{FREYA}$ simulation 
based on microscopic distributions
that can be directly compared
with experimental data,
ranging between $N_{\gamma} = 6.88 \pm 0.35$ \cite{pleasonton1973} and
$N_{\gamma} = 7.23 \pm 0.3$ \cite{verbinski1973}. 

\begin{table}[!ht]
\caption{\label{tab:energies} Total average photon multiplicities $N_{\gamma}$ for several 
fragmentations in $^{239}$Pu($n_{\text{th}}$,f) calculated with $\mathtt{FREYA}$,
based on the microscopic DFT or default $\mathtt{FREYA}$ AM
distributions. Mass 
yields $Y(A_H)$ and charge yields $Y(Z_H)$
(each normalized to $100$ for the heavy fragment only) 
used in $\mathtt{FREYA}$ \cite{verbeke2018} give an estimates 
of the relative importance of each fragmentation.}
\vspace{2mm}
\begin{tabular}{cccccc} 
{($Z_H,A_H$)} & {$Y(Z_H)$} & {$Y(A_H)$}  & 
{$N_{\gamma}({\rm{DFT}})$} & {$N_{\gamma}({\rm Default})$}  \vspace*{1mm} \\ 
\hline \hline
$(50, 128)$  & $2.7$ & $1.04$ & $11.72$ & $8.69$ \\
$(50, 130)$  & $2.7$ & $2.12$  & $6.35$  & $9.65$ \\
$(50, 132)$  & $2.7$ & $3.70$ & $4.89$  & $6.62$ \\
$(54, 138)$  & $15.3$ & $6.09$ & $7.53$  & $7.32$ \\
$(56, 144)$  & $12.1$ & $4.38$ & $12.46$  & $7.41$ \\ 
\end{tabular} 
\label{tab:multiplicities}
\end{table} 

Finally, the average angular momenta of FFs in the neutron-induced fission 
of $^{232}$Th and $^{238}$U and 
spontaneous fission of $^{252}$Cf were very recently measured at the ALTO 
facility in Orsay \cite{wilson2021}. 
The authors obtained angular momenta of the FFs after 
the emission of neutrons and statistical photons 
(hereafter referred to as the post-emission AM). They inferred 
the FF angular momenta prior to statistical photon emission by adding 
a constant $1 \hbar$ shift, according to the prescription 
of Ref.~\cite{abdelrahman1987}.
The post-emission AM in all three reactions are in the 
$J_{H, L} \approx 3-7 \hbar$ range and exhibit a sawtooth mass 
dependence which is proposed to be universal. 
Our calculated post-emission AM in $^{239}$Pu($n_{\text{th}}$,f) 
(Fig.~\ref{fig:average_AM}b) are in 
a very similar range, but the impact of statistical photons is 
markedly larger and more 
complex than the constant $1 \hbar$ shift adopted in \cite{wilson2021}.
The $J_H (A_H)$ values calculated both 
before and after the emission of statistical photons are consistent 
with the proposed linear mass dependence.
On the other hand,
while the post-emission $J_L(A_L)$ values are not incompatible with
a weak linear dependence, they appear more consistent with 
a horizontal line. 
The overall pattern we observe
is consistent with the proposed universal sawtooth pattern.
In addition, we stress that we typically do observe 
asymmetries in the AM of partner primary
fragments. In particular,
a light fragment typically carries
more AM than its
heavy counterpart (for example,
$7.3(0.1) \hbar$ in $\ce{^{110}_{44}}$Ru and
$2.8(0.1) \hbar$ in $\ce{^{130}_{50}}$Sn), but
the opposite is also possible 
($13.7(0.2) \hbar$ in $\ce{^{150}_{58}}$Ce and $11.9(0.3) \hbar$ in $\ce{^{90}_{36}}$Kr). 
This invalidates the claim from \cite{wilson2021} that models based on the angular momentum
generation at scission due to the deformation
of FFs
cannot account for such asymmetries. 
%A further refinement of theoretical modeling 
%by carrying out the PNP in FFs, including the finite temperature effects, 
%and performing a more advanced uncertainty quantification
%will help bridge the gap between the theory
%and the experiment and shed more light
%on the microscopic origin of 
%angular momentum in fission fragments.

\indent{\it Conclusion. ---} We have presented
the first microscopic calculation of angular momentum distributions for a
wide range of fission fragment masses. These calculations reveal the large 
impact of the underlying
shell structure of the fission fragments at scission on their angular momentum. 
The average angular momenta of the primary fission fragments
exhibit a 
marked dip near the doubly magic fragment $^{132}$Sn
and the resulting mass 
dependence is reminiscent of the
recently observed sawtooth pattern \cite{wilson2021}. The present work demonstrates
the importance of using a predictive 
theory to compute properties that can vary significantly from one fragmentation
to the next and which could thus provide valuable guidance for nuclear data evaluation. 
This novel treatment may also be useful for describing the properties of fragments
in very neutron-rich nuclei, such as those involved in nucleosynthesis
processes, where experimental data are unavailable.

%%%%%%%%%%%%%%%%%%%%%%%%%%%%%%%%%%%%%%%%%%%%%%%%%%%%%%%%%%%%%%%%%%%%%%%%%%%%%%%
%%%%%%%%%%%%%%%%%%%%%%%%%%%%%%%%%%%%%%%%%%%%%%%%%%%%%%%%%%%%%%%%%%%%%%%%%%%%%%%
%%%%%%%%%%%%%%%%%%%%%%%%%%%%%%%%%%%%%%%%%%%%%%%%%%%%%%%%%%%%%%%%%%%%%%%%%%%%%%%
%%%%%%%%%%%%%%%%%%%%%%%%%%%%%%%%%%%%%%%%%%%%%%%%%%%%%%%%%%%%%%%%%%%%%%%%%%%%%%%

\indent{\it Acknowledgements. ---} P. M. and N. S. acknowledge stimulating discussions with
Marc Verri\`ere and David Regnier.
R. V. thanks Robert V. F. Janssens for discussion.
This work was performed under the auspices of the U.S.\ 
Department of Energy by Lawrence Livermore National Laboratory under Contract 
DE-AC52-07NA27344 and under the NUCLEI SciDAC-4 collaboration. 
J. R. acknowledges support from the Office of Nuclear Physics 
in the U.S. Department of Energy under Contracts DE-AC02-05CH11231. Computing support for this work came from the Lawrence 
Livermore National Laboratory (LLNL) Institutional Computing Grand Challenge
program.

%%%%%%%%%%%%%%%%%%%%%%%%%%%%%%%%%%%%%%%%%%%%%%%%%%%%%%%%%%%%%%%%%%%%%%%%%%%%%%%
%%%%%%%%%%%%%%%%%%%%%%%%%%%%%%%%%%%%%%%%%%%%%%%%%%%%%%%%%%%%%%%%%%%%%%%%%%%%%%%
%%%%%%%%%%%%%%%%%%%%%%%%%%%%%%%%%%%%%%%%%%%%%%%%%%%%%%%%%%%%%%%%%%%%%%%%%%%%%%%
%%%%%%%%%%%%%%%%%%%%%%%%%%%%%%%%%%%%%%%%%%%%%%%%%%%%%%%%%%%%%%%%%%%%%%%%%%%%%%%

\bibliographystyle{unsrt}
%\bibliography{zotero_output,books}
\bibliography{AngularMomentumOfFragments}

\end{document}